\documentclass[journal]{IEEEtran}

\usepackage{amsmath}
\usepackage{amssymb}
\usepackage{amsfonts}
\usepackage{graphicx}
\usepackage{epsfig}
\usepackage{subfigure}
\usepackage{psfrag}
\usepackage{cite}
\usepackage{latexsym}
\usepackage{url}
\usepackage{color}
\usepackage{multirow}

\PassOptionsToPackage{bookmarks={false}}{hyperref}

\begin{document}
\title{Fundamental Rate Limits of Physical Layer Spoofing
%\author{Jie Xu$^1$, Lingjie Duan$^1$, and Rui Zhang$^2$\\
%$^1$Engineering Systems and Design Pillar, Singapore University of Technology and Design\\
%$^2$Department of Electrical and Computer Engineering, National University of Singapore\\
%E-mail:~\{jie\_xu,~lingjie\_duan\}@sutd.edu.sg,~elezhang@nus.edu.sg}
}
%\author{Jie Xu$^1$, Lingjie Duan$^1$, and Rui Zhang$^2$\\
%\author{Jie Xu, Lingjie Duan, and Rui Zhang\\
\author{Jie Xu, Lingjie Duan, and Rui Zhang\\
%\thanks{Part of this paper will be presented in the IEEE Global Communications Conference (GLOBECOM), Washington, DC USA, December 4-8, 2016.}
\thanks{J. Xu is with the School of Information Engineering, Guangdong University of Technology (e-mail: jiexu.ustc@gmail.com). He is also with the Engineering Systems and Design Pillar, Singapore University of Technology and Design.}
\thanks{L. Duan is with the Engineering Systems and Design Pillar, Singapore University of Technology and Design (e-mail:~lingjie\_duan@sutd.edu.sg).}
\thanks{R. Zhang is with the Department of Electrical and Computer Engineering, National University of Singapore (e-mail: elezhang@nus.edu.sg). He is also with the Institute for Infocomm Research, A*STAR, Singapore.}
}

\maketitle

\begin{abstract}
This letter studies an emerging wireless communication intervention problem at the physical layer, where a legitimate spoofer aims to spoof a malicious link from Alice to Bob, by replacing Alice's transmitted source message with its target message at Bob side. From an information-theoretic perspective, we are interested in characterizing the maximum achievable spoofing rate of this new spoofing channel, which is equivalent to the maximum achievable rate of the target message at Bob, under the condition that Bob cannot decode the source message from Alice. We propose a novel combined spoofing approach, where the spoofer sends its own target message, combined with a processed version of the source message to cancel the source message at Bob. For both cases when Bob treats interference as noise (TIN) or applies successive interference cancelation (SIC), we obtain the maximum achievable spoofing rates by optimizing the power allocation between the target and source messages at the spoofer.\vspace{-0em}
\end{abstract}
\begin{keywords}
Wireless communication intervention, physical layer spoofing, achievable spoofing rate, power allocation.\vspace{-1em}
\end{keywords}

\newtheorem{definition}{\underline{Definition}}[section]
\newtheorem{fact}{Fact}
\newtheorem{assumption}{Assumption}
\newtheorem{theorem}{\underline{Theorem}}[section]
\newtheorem{lemma}{\underline{Lemma}}[section]
\newtheorem{corollary}{\underline{Corollary}}[section]
\newtheorem{proposition}{\underline{Proposition}}[section]
\newtheorem{example}{\underline{Example}}[section]
\newtheorem{remark}{\underline{Remark}}[section]
\newtheorem{algorithm}{\underline{Algorithm}}[section]
\newcommand{\mv}[1]{\mbox{\boldmath{$ #1 $}}}
\setlength\abovedisplayskip{3pt}
\setlength\belowdisplayskip{3pt}

\section{Introduction}

The emergence of infrastructure-free wireless communications (e.g., mobile ad hoc networks and unmanned aerial vehicle (UAV) communications) imposes new challenges on the public security, since they may be misused by malicious users to commit crimes or even terror attacks \cite{Mag}. To overcome this issue, authorized parties can launch legitimate information eavesdropping (see, e.g., \cite{XuDuanZhang1,XuDuanZhang2,ZengZhang,ZengZhang2}) and jamming (see, e.g., \cite{ZouWangHanzo2015,Medard,Kashyap2004,Liu2015}) on suspicious and malicious wireless communication links, so as to monitor and intervene in them for the purpose of detecting and preventing security attacks \cite{Mag}.

We focus on the emerging wireless communication intervention at the physical layer. Different from the jamming intervention that can only disrupt or disable target links, we propose a new intervention via physical layer spoofing to change the communicated information over malicious links while keeping their operation. Such a physical layer spoofing has been first investigated in our previous work \cite{Xu2016a} by considering a three-node spoofing channel (see Fig. \ref{fig:1}), where a legitimate spoofer aims to spoof an ongoing malicious link from Alice to Bob, by replacing Alice's transmitted source message with its target message at Bob side. We have proposed a symbol-level spoofing approach in \cite{Xu2016a} for the spoofer to minimize the spoofing-symbol-error-rate of the target message at Bob under practical phase-shift keying modulations. Nevertheless, the fundamental information-theoretic limits of such a new spoofing channel remain unaddressed, thus motivating our study in this work.

In this letter, we are interested in characterizing the maximum achievable spoofing rate of the spoofing channel in Fig. \ref{fig:1}, which is equivalent to the maximum achievable rate of the target message at Bob, while ensuring that Bob cannot decode the source message. We propose a new combined spoofing approach, where the spoofer sends its own target message, combined with a processed version of the source message to cancel it at Bob. In particular, we assume Alice transmits with a constant rate, and consider two cases when Bob treats interference as noise (TIN) and applies successive interference cancelation (SIC), respectively. To successfully spoof in the former case, the spoofer should make the received target message at Bob be stronger than the source message; and in the latter case, the spoofer should make the maximum achievable rate of the source message (under different decoding orders) be strictly smaller than Alice's communication rate. In both cases, we obtain the maximum achievable spoofing rates by optimizing the power allocation between the target and source messages at the spoofer. Numerical results show that our proposed combined spoofing approach with optimized power allocation outperforms other benchmark spoofing schemes.\vspace{-1.2em}

\section{System Model}

\begin{figure}
\centering
 \epsfxsize=1\linewidth
    \includegraphics[width=6.5cm]{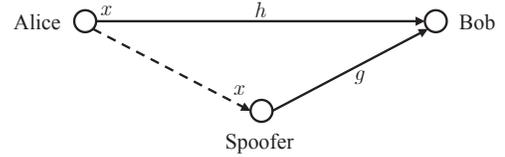}
\caption{A three-node spoofing channel, where a legitimate spoofer aims to change the communicated information from Alice to Bob.} \label{fig:1}\vspace{-2em}
\vspace{-0cm}
\end{figure}

As shown in Fig. \ref{fig:1}, we consider a three-node spoofing channel, where two malicious users Alice and Bob communicate to plan or commit crimes, and a legitimate spoofer aims to change the communicated data from Alice to Bob to defend against them. Practically, the malicious users can be identified {\it a priori} via, e.g., legitimate information eavesdropping \cite{XuDuanZhang1,XuDuanZhang2,ZengZhang,ZengZhang2}. We define $h$ and $g$ as the complex channel coefficients of the malicious link from Alice to Bob and the spoofing link from the spoofer to Bob, respectively.

First, we consider the case without spoofing. Let $s$ denote the source message transmitted by Alice with unit power.  The received signal by Bob is given by
\begin{align}\label{eqn:system:model}
y = h \sqrt{P} s + n,
\end{align}
where $P$ denotes the constant transmit power of Alice, and $n$ denotes the receiver noise at Bob being a circularly symmetric complex Gaussian (CSCG) random variable with zero mean and unit variance. The capacity of the malicious link is given as $C \triangleq \log_2(1 + |h|^2P)$, which is achieved when Alice employs Gaussian signaling (i.e., setting $s$ as a CSCG random variable with zero mean and unit variance). Suppose that Alice communicates with Bob with a constant communication rate $R$ no greater than the channel capacity $C$, i.e., $R \le C$, where $R$ is chosen based on the quality of service (QoS) requirement.

Next, we consider the case with spoofing. The spoofer aims to change Bob's decoded message from Alice's source message $s$ to its desired target message. It is assumed that the spoofer has the perfect information of the source message $s$ and the channel coefficients $h$ and $g$. This assumption is made to help derive the spoofing rate upper bound, similar to that in the prior works in the information-theoretic literature (see, e.g., the correlated jamming in \cite{Kashyap2004} and the cognitive radio channel in \cite{Devroye2006}).{\footnote{Though beyond the scope of this letter, please refer to \cite{Xu2016a} for a detailed example for the spoofer to practically obtain $s$, $h$ and $g$, and synchronize with Alice and Bob, where the spoofer can act as a fake relay in the malicious network in obtaining such information. Furthermore, the spoofer can work in a full-duplex mode (e.g., amplify and forward) to obtain $s$ via eavesdropping from Alice and at the same time spoof Bob.}}  In this case, the spoofer can use the same codebook of $s$ for sending the target message, such that Bob will decode the target message without awareness of being spoofed. Let $x$ denote the target message with unit power, which is in general independent of $s$. We consider a combined spoofing approach, where the spoofer designs its spoofing signal $z$ to be a combined version of both the source message $s$ and the target message $x$ with proper processing. Particularly, we have $z = \alpha s + \beta x$, where $\alpha$ and $\beta$ denote the complex transmit coefficients for the messages $s$ and $x$, respectively. In this case, the received signal $y$ at Bob can be expressed as
\begin{align}\label{eqn:system:model:2}
y = h\sqrt{P} s + g z + n = \left(h\sqrt{P} + g \alpha \right)s + g \beta x + n.
\end{align}
By denoting $Q$ as the maximum spoofing power at the spoofer, then we have
\begin{align}\label{eqn:sum_power}
|\alpha|^2 + |\beta|^2 \le Q.
\end{align}

In order to successfully spoof the malicious communication, the spoofer should design the spoofing signal (i.e., the transmit coefficients $\alpha$ and $\beta$) such that Bob is only able to successfully decode the target message $x$ but fails to decode the source message $s$. In this case, the successful spoofing critically depends on the decoding method employed by Bob. We consider two typical Bob receivers as follows, including the practical TIN receiver and the information-theoretically optimal SIC receiver. It is assumed that the spoofer is aware of which receiver being employed by Bob.

\subsubsection{TIN receiver at Bob}

Bob does not know the coexistence of the two messages $s$ and $x$, and thus considers the stronger one between them to be its desired signal, and treats the other one (the co-channel interference) to be noise. In this case, the received message $x$ at Bob should have a stronger power than $s$ such that the spoofing is successful.

\subsubsection{SIC receiver at Bob}

Bob is able to detect the coexistence of $s$ and $x$, and accordingly attempts to use SIC to decode both of them. From the successfully decoded ones (if any), Bob will decide which  the desired message is. In particular, Bob first decodes one message ($x$ or $s$) by treating the other as noise, and then cancels it from the received message $y$ to decode the other one. Generally speaking, Bob can use two different decoding orders (first $x$ and then $s$, or first $s$ and then $x$).%, and accordingly the spoofer needs to ensure Bob unable to decode $s$ under both orders.

Under both receiver cases, we aim to characterize the maximum achievable spoofing rates of the target message $x$, provided that Bob cannot decode the source message $s$.{\footnote{In practice, the spoofer can choose any rate (for $x$) no larger than the maximum achievable spoofing rate, provided with successful spoofing.}} \vspace{-1.5em}

\section{Spoofing TIN Receiver at Bob}

\subsection{Problem Formulation for TIN Receiver}

When Bob employs the TIN receiver, the spoofer can successfully spoof the malicious communication link only when the received power of the target message $x$ is greater than that of the source message $s$. Mathematically, it must hold that $|g\beta|^2 > |h\sqrt{P} + g \alpha|^2$. Note that this strict inequality constraint may make the associated optimization problem ill-posed: an optimizer on the boundary of the feasible region may not be attainable. To address this issue, we revise it to be a non-strict inequality constraint as
\begin{align}
|g\beta|^2 \ge |h\sqrt{P} + g \alpha|^2 + \delta_1 ,\label{eqn:successful:spoofing:nonstrict}
\end{align}
where $\delta_1 > 0$ is a sufficiently small positive constant.

In this case, the received signal-to-interference-plus-noise-ratio (SINR) for the target message $x$ at Bob is
$\gamma (\alpha,\beta)= \frac{|g\beta|^2}{|h\sqrt{P} + g \alpha|^2 + 1}$. Accordingly, the achievable spoofing rate (in bps/Hz) is expressed as follows by assuming $x$ is CSCG and $s$ is also CSCG as the ``worst-case'' noise.
\begin{align}
r(\alpha,\beta) = \log_2\left( 1+ \frac{|g\beta|^2}{|h\sqrt{P} + g \alpha|^2 + 1} \right).\label{eqn:r:alpha}
\end{align}
As a result, the achievable spoofing rate maximization problem is formulated as
\begin{align}
\mathrm{(P1)}:~\max_{\alpha,\beta} ~& r (\alpha,\beta) \nonumber\\
\mathrm{s.t.}~& (\ref{eqn:sum_power})~{\rm and}~(\ref{eqn:successful:spoofing:nonstrict}).\nonumber
\end{align}
\vspace{-2em}

\subsection{Optimal Spoofing Solution to Problem (P1)}

First, we reformulate (P1) as an equivalent problem with a single real decision variable. It is evident that the optimality of (P1) is attained when the processed source message $s$ from the spoofer is destructively combined at Bob with that from Alice, and the sum-power constraint in (\ref{eqn:sum_power}) is tight. In other words, we have
\begin{align}
\alpha =& -\frac{hg^*}{|h||g|}\tilde\alpha,\label{eqn:alpha:opt}\\
\beta =& \sqrt{Q - |\tilde\alpha|^2},\label{eqn:beta:opt}
\end{align}
where the superscript $*$ denotes the conjugate operation, and $\tilde\alpha \ge 0$ denotes the magnitude of $\alpha$. Here, since both the objective function and constraints of (P1) are irrespective of the phase of $\beta$, in (\ref{eqn:beta:opt}) we decide $\beta$ to be a real variable without loss of optimality. %With (\ref{eqn:success:1}) replaced by (\ref{eqn:successful:spoofing:nonstrict}) and by substituting (\ref{eqn:alpha:opt}) and (\ref{eqn:beta:opt})
Therefore, (P1) is equivalently reformulated as follows to optimize an SINR function $\tilde\gamma(\tilde\alpha)$ with only a real decision variable $\tilde\alpha$.
\begin{align}
&\mathrm{(P1.1)}:\max_{\tilde\alpha \ge 0} ~ \tilde\gamma(\tilde\alpha) \triangleq \frac{|g|^2(Q-\tilde\alpha^2)}{(|h|\sqrt{P} - |g|\tilde\alpha)^2 + 1}   \nonumber\\
&\mathrm{s.t.} ~2|g|^2\tilde\alpha^2 - 2|h||g|\sqrt{P} \tilde\alpha+ |h|^2P - |g|^2Q + \delta_1 \le 0.\label{eqn:successful:spoofing:reform}
\end{align}

Next, we check the feasibility of problem (P1.1) (and thus (P1)).
\begin{lemma}\label{lemma1}
Problem (P1.1) (and thus (P1)) is feasible if and only if $Q \ge \frac{|h|^2P + 2\delta_1}{2|g|^2}$.
\end{lemma}
\begin{IEEEproof}
Note that the constraint in (\ref{eqn:successful:spoofing:reform}) can be rewritten as $2|g|^2\left(\tilde\alpha - \frac{|h|\sqrt{P}}{2|g|}\right)^2 + \frac{|h|^2P}{2} - |g|^2Q + \delta_1 \le 0$, which specifies a nonempty feasible set if and only if $\frac{|h|^2P}{2} - |g|^2Q + \delta_1 \le 0$. Equivalently, problem (P1.1) is feasible if and only if $Q \ge \frac{|h|^2P + 2\delta_1}{2|g|^2}$. This proposition thus follows.
\end{IEEEproof}

Finally, we obtain the optimal solutions to (P1.1) and (P1) when they are feasible. In this case, the constraint in (\ref{eqn:successful:spoofing:reform}) is equivalently expressed as
\begin{align}\label{eqn:feasible}
{\underline{\omega}} \le \tilde\alpha \le {\overline{\omega}},
\end{align}
where ${\underline{\omega}} = \frac{|h|\sqrt{P} - \sqrt{2|g|^2Q - |h|^2P - 2\delta_1}}{2|g|}$ and ${\overline{\omega}} = \frac{|h|\sqrt{P} + \sqrt{2|g|^2Q - |h|^2P - 2\delta_1}}{2|g|}$ denote the minimum and maximum values of $\tilde\alpha$ for the TIN spoofing to be successful, respectively. Furthermore, by checking its first-order derivative, we can show that there exist one local maximum point $\tilde\alpha_1$ and one local minimum point $\tilde\alpha_2$ for the SINR function $\tilde\gamma(\tilde\alpha)$, which are given by \vspace{-1em}

\begin{small}\begin{align}\label{eqn:alpha1}
\tilde\alpha_1 =  \frac{|h|^2P + |g|^2Q + 1}{2|h||g|\sqrt{P}} - \frac{ \sqrt{\left(|h|^2P + |g|^2Q + 1\right)^2 - 4 |h|^2|g|^2 PQ}}{2|h||g|\sqrt{P}}
\end{align}\end{small}and $\tilde\alpha_2 =  \frac{|h|^2P + |g|^2Q + 1}{2|h||g|\sqrt{P}} + \frac{ \sqrt{\left(|h|^2P + |g|^2Q + 1\right)^2 - 4 |h|^2|g|^2 PQ}}{2|h||g|\sqrt{P}}$, respectively. In particular, $\tilde\gamma(\tilde\alpha)$ is first increasing over $\tilde\alpha \in [0,\tilde\alpha_1]$, then decreasing over $\tilde\alpha \in (\tilde\alpha_1,\tilde\alpha_2)$, and finally increasing over  $\bar\alpha \in [\tilde\alpha_2,+\infty)$. Since $\lim_{\tilde\alpha \to \infty} \tilde\gamma(\tilde\alpha) = -1$ but $\tilde\gamma(\tilde\alpha) > 0, \forall \tilde\alpha \in [0,\tilde\alpha_1]$, it is evident that $\tilde\alpha_1$ is the globally optimal point to maximize $\tilde\gamma(\tilde\alpha)$ without any constraints. Then we have the following proposition.
\begin{proposition}\label{proposition1}
The optimal solution to (P1.1) is given by
\begin{align}\label{eqn:solution:P1.1}
\tilde\alpha^{\star} = \max\left({\underline{\omega}},\tilde\alpha_1\right),
\end{align}
and thus the optimal solution to (P1) is $\alpha^{\star} = -\frac{hg^*}{|h||g|}\tilde\alpha^\star$ and $\beta^\star = \sqrt{Q - |\alpha^{\star}|^2}$.
\end{proposition}

\begin{IEEEproof}
The optimal solution to (P1.1) can be easily verified based on the monotonic property of $\tilde\gamma(\tilde\alpha)$ together with the fact that $\tilde\alpha_1 \le {\overline{\omega}}$ and $\tilde\alpha_1 \le \sqrt{Q}$. Then, by substituting $\tilde\alpha^{\star}$ in (\ref{eqn:solution:P1.1}) into (\ref{eqn:alpha:opt}) and (\ref{eqn:beta:opt}), the optimal solution to (P1) is derived. Therefore, this proposition is proved.
\end{IEEEproof}

\vspace{-0em}

\section{Spoofing SIC Receiver at Bob}

\subsection{Problem Formulation for SIC Receiver}

When Bob employs the SIC receiver, the spoofer needs to design its spoofing signal such that Bob is able to decode the target message $x$ but fails to decode the source message $s$ for the purpose of successful spoofing. In general, the spoofer should consider the following two cases, depending on the decoding orders employed by Bob. Here, Bob can be viewed as a receiver of a two-user multiple-access channel (MAC) by considering Alice and the spoofer as the two transmitters.

In the first case, Bob first decodes $s$ by treating $x$ as noise, and then subtracts $s$ from the received signal $y$ to decode $x$. Accordingly, the maximum achievable rates of $s$ and $x$ at the receiver of Bob (under given $\alpha$ and $\beta$) are given as follows by assuming both $s$ and $x$ are CSCG. % \cite[Chapter 6]{Tse}
    \begin{align}
    r_s^{({\rm I})} & = \log_2\left(1+ \frac{|h \sqrt{P} + g \alpha|^2}{|g\beta|^2 + 1}\right),\\
    r_x^{({\rm I})} & = \log_2\left(1+ |g\beta|^2\right).\label{eqn:r_2:1}
    \end{align}
In order to prevent Bob from successfully decoding $s$, the spoofer should ensure that its maximum achievable rate is smaller than Alice's communication rate, i.e.,
\begin{align}\label{eqn:case1}
r_s^{({\rm I})} < R.
\end{align}
Since Bob fails to decode $s$, the decoding of $x$ should suffer from the interference of $s$, and therefore the achievable spoofing rate is given as $r(\alpha,\beta)$ in (\ref{eqn:r:alpha}).

In the second case, Bob first decodes $x$ by treating $s$ as noise, and then cancels $x$ from $y$ to decode $s$. Accordingly, the maximum achievable rates of $s$ and $x$ (under given $\alpha$ and $\beta$) at the receiver of Bob are respectively given by
    \begin{align}
    r_s^{({\rm II})} & = \log_2\left(1+ |h \sqrt{P} + g \alpha|^2 \right),\\
    r_x^{({\rm II})} & = \log_2\left(1+ \frac{|g\beta|^2}{|h \sqrt{P} + g \alpha|^2 + 1}\right).\label{eqn:r_2:2}
    \end{align}
To prevent Bob from decoding $s$, the spoofer should ensure that
\begin{align}\label{eqn:case2}
r_s^{({\rm II})}  < R.
\end{align}
In this case, the rate $r_x^{({\rm II})}$ in (\ref{eqn:r_2:2}), which equals $r(\alpha,\beta)$ in (\ref{eqn:r:alpha}), is the achievable spoofing rate.

By combining the two cases, the successful spoofing only requires (\ref{eqn:case2}) to hold, since if it holds, (\ref{eqn:case1}) will hold automatically. Note that in the above two cases, Bob cannot decode $s$ regardless of the decoding orders used with SIC; as a result, it can only treat the decoded target message $x$ as its desired message. Also note that (\ref{eqn:case2}) is a strict inequality constraint. To address this issue, we revise (\ref{eqn:case2}) as follows similarly as in (\ref{eqn:successful:spoofing:nonstrict}).
\begin{align}
\log_2(1+|h\sqrt{P} + g \alpha|^2) + \delta_2 \le R,\label{eqn:feasible:case2}
\end{align}
where $\delta_2$ is a sufficiently small positive constant. The achievable spoofing rate maximization problem is formulated as
\begin{align}
\mathrm{(P2)}:~\max_{\alpha,\beta} ~&r(\alpha,\beta)\nonumber\\
\mathrm{s.t.}~&(\ref{eqn:sum_power})~{\rm and}~(\ref{eqn:feasible:case2}).\nonumber
\end{align}\vspace{-2em}

\vspace{-1em}
\subsection{Optimal Spoofing Solution to Problem (P2)}

Similar to (P1), it can be shown that the optimality of (P2) is attained when (\ref{eqn:alpha:opt}) and (\ref{eqn:beta:opt}) hold. In this case, (P2) is equivalently reformulated as
\begin{align}
\mathrm{(P2.1)}:~\max_{0\le \tilde\alpha \le \sqrt{Q}} ~& \tilde\gamma (\tilde\alpha)   \nonumber\\
\mathrm{s.t.}~&  \underline{\chi} \le \tilde\alpha \le \overline{\chi},\label{eqn:feasibility:reform:2}
\end{align}
where $\underline{\chi} = \frac{|h|\sqrt{P} - \sqrt{2^{R-\delta_2} - 1}}{|g|}$ and $\overline{\chi} =  \frac{|h|\sqrt{P} + \sqrt{2^{R-\delta_2} - 1}}{|g|}$ denote the minimum and maximum values of $\tilde\alpha$ for the SIC spoofing to be successful, respectively.

Next, we check the feasibility of problem (P2.1) (and thus (P2)).
\begin{lemma}\label{lemma2}
Problem (P2.1) (and thus (P2)) is feasible if and only if $Q \ge \underline{\chi}^2$.
\end{lemma}
\begin{IEEEproof}
The feasible condition of problem (P2.1) can be obtained by noting that $\tilde\alpha \le \sqrt{Q}$ and $\underline{\chi} \le \tilde\alpha$ should be satisfied at the same time.
\end{IEEEproof}

Finally, when problems (P2.1) and (P2) are feasible, their optimal solutions are obtained in the following proposition.
\begin{proposition}\label{proposition2}
The optimal solution to (P2.1) is given by
\begin{align}\label{eqn:propo2:eq1}
&\bar\alpha^{\star\star} = \max\left( \underline{\chi} ,\tilde\alpha_1\right),
%\nonumber\\
%&\left\{
%\begin{array}{ll}
%\bar\alpha_2,&{\rm if}~ \frac{|h|\sqrt{P} - \sqrt{2^{R-\delta_2} - 1}}{|g|} \le \bar\alpha_2 \le \sqrt{Q},\\
%\frac{|h|\sqrt{P} -  \sqrt{2^{R-\delta_2} - 1}}{|g|},&{\rm if}~\frac{|h|\sqrt{P} -  \sqrt{2^{R-\delta_2} - 1}}{|g|} > \bar\alpha_2,\\
%Q,&{\rm if}~\bar\alpha_2 > \sqrt{Q}.
%\end{array}
%\right.
\end{align}
where  $\tilde\alpha_1$ is the globally optimal point to maximize $\tilde\gamma(\tilde\alpha)$, as given in (\ref{eqn:alpha1}). Then, the optimal solution to (P2) is given by $\alpha^{\star\star} = -\frac{hg^*}{|h||g|}\bar\alpha^{\star\star}$ and $\beta^{\star\star} = \sqrt{Q - |\alpha^{\star\star}|^2}$.
\end{proposition}
\begin{IEEEproof}
Similar to Proposition \ref{proposition1} and based on the monotonic property of $\tilde\gamma(\tilde \alpha)$, the optimal solution to (P2.1) is obtained as $\bar\alpha^{\star\star}$ in (\ref{eqn:propo2:eq1}). Substituting it into (\ref{eqn:alpha:opt}) and (\ref{eqn:beta:opt}), the optimal solution to (P2) is derived. Therefore, this proposition is proved.
 \end{IEEEproof}

It is interesting to compare the optimally designed spoofing signals for TIN and SIC receivers at Bob, respectively. First, it is observed from Lemmas \ref{lemma1} and \ref{lemma2} that the minimally required spoofing power for the TIN receiver is irrespective of the communication rate $R$ by Alice, while that for the SIC receiver is monotonically decreasing with respect to $R$. As a result, when $R$ is large (particularly when $2^R-1 > (1-\frac{1}{\sqrt{2}})^2|h|^2{P}$ by neglecting the sufficiently small $\delta_1$ and $\delta_2$), the minimally required spoofing power for the SIC receiver is smaller than that for the TIN receiver, and thus the SIC Bob receiver is easier to be spoofed than the TIN one in this case. Next, it is observed from Propositions  \ref{proposition1} and \ref{proposition2} that if $\tilde\alpha_1 \ge \underline{\omega}$ and $\tilde\alpha_1 \ge \underline{\chi}$ both hold, then the designed spoofing signals become identical for both receivers. This happens when the spoofing power budget $Q$ becomes sufficiently large.\vspace{-0em}

\vspace{-0.5em}
%\section{Optimal Spoofing Design}

\section{Numerical Results}
\vspace{-0em}
In this section, we provide numerical results to show the achievable spoofing rates of our proposed combined spoofing approach with optimal spoofing signals design. We compare our results with two benchmark schemes in the following.
\begin{itemize}
 \item {\it Heuristic combined spoofing with perfect source message cancelation}: The spoofer tries to cancel all the source message by setting $\alpha = -\frac{hg^*\sqrt{P}}{|g|^2}$, and accordingly $\beta$ is given in (\ref{eqn:beta:opt}). This scheme only works when $Q>\frac{|h|^2P}{|g|^2}$ for both TIN and SIC Bob receivers, where the minimally required spoofing power $\frac{|h|^2P}{|g|^2}$ is twice of that in Lemma \ref{lemma1} for our proposed optimal combined spoofing.
 \item {\it Naive spoofing}: The spoofer uses all its transmit power to send the target message $s$, which corresponds to the case with $\alpha = 0$ and $\beta = \sqrt{Q}$. This scheme only applies to the case with the TIN receiver at Bob when $Q>\frac{|h|^2P}{|g|^2}$.
\end{itemize}

\begin{figure}
\centering
 \epsfxsize=1\linewidth
    \includegraphics[width=7cm]{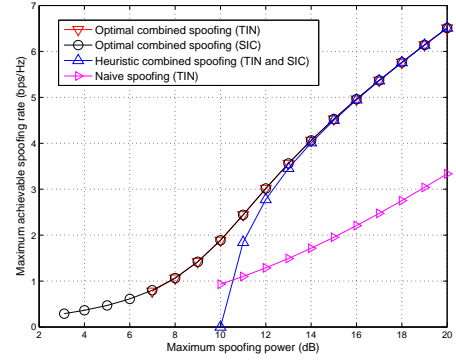}
\caption{The maximum achievable spoofing rate versus the spoofing power $Q$ at the spoofer.} \label{fig:2}
\vspace{-1.5em}
\end{figure}

In the simulation, we normalize the channel coefficients to be $h=1$ and $g=1$ for the purpose of illustration, while our results can be easily extended to the other values of $h$ and $g$. We set $P=10$ dB, and $R = 2$ bps/Hz. Fig. \ref{fig:2} shows the maximum achievable spoofing rate versus the spoofing power $Q$ at the spoofer. It is observed that the two benchmark schemes achieve positive spoofing rates (or successfully spoof) only when $Q > 10$ dB, while the optimal combined spoofing does so when $Q > 5$ dB for the TIN receiver at Bob and when $Q$ is larger than 3 dB for the SIC receiver. It is also observed that when $Q$ is larger than $7$ dB, the optimal combined spoofing achieves the same maximum achievable spoofing rate for both TIN and SIC receivers, and outperforms both benchmarks schemes. The heuristic combined spoofing is observed to achieve the same performance as the optimal one when $Q > 16$ dB. This shows that in this case, it is optimal for the spoofer to perfectly cancel the source message and then allocate the remaining power for the target message.\vspace{-0em}
\vspace{-0.5em}

\section{Conclusion}\vspace{-0em}

This letter studied the achievable spoofing rates of the new wireless communication intervention via physical layer spoofing, where a legitimate spoofer sends a combined version of both the source and target messages to confuse a malicious link from Alice to Bob. We proposed optimal spoofing signal designs when Bob employs the TIN and SIC receivers, respectively. It is our hope that this work can provide new insights on the fundamental information-theoretic limits of the physical layer spoofing. How to extend the results to general multi-antenna and multiuser scenarios is an interesting research direction worth pursuing in the future work.

\end{document}